\documentclass[epj]{svjour}
\usepackage{graphics}
\usepackage{graphicx}
\usepackage{amsmath}
\usepackage{psfrag}
\usepackage{subfigure}
\usepackage{amsfonts}

\newcommand{\ket}[1]{|#1\rangle} 
\newcommand{\bra}[1]{\langle#1|}

\newcommand{\erwwert}[1]{\langle#1\rangle}
\newcommand{\Sp}[2][]{\text{Tr}_{#1}\{#2\}}
\newcommand{\ee}{\text{e}}                 

\begin{document}
\title{Relaxation into equilibrium under pure Schr\"odinger dynamics}
\author{P. Borowski\inst{1}\thanks{borowski@theo1.physik.uni-stuttgart.de} \and J. Gemmer\inst{1} \and G. Mahler\inst{1}
}                     
\offprints{}          
\institute{Institut f\"ur Theoretische Physik I - Universit\"at Stuttgart, Pfaffenwaldring 57, 70550 Stuttgart, Germany}
\date{Received: date / Revised version: date}
%
\abstract{
We consider bipartite quantum systems that are described completely by a state vector $\ket{\Psi(t)}$ and the fully deterministic Schr\"odinger equation. Under weak constraints and without any artificially introduced decoherence or irreversibility, the smaller of the two subsystems shows thermodynamic behaviour like relaxation into an equilibrium, maximization of entropy and the emergence of the Boltzmann energy distribution. This generic behaviour results from entanglement.
}
\maketitle
\section{Introduction}
\label{sec_introduction}
Various attempts have been made for quantum systems to account for thermodynamic behaviour~\cite{Landau_Lifshitz}~\cite{von-Neumann}~\cite{Lindblad}, in particular for the relaxation into an equilibrium state. Analogous to the Hamilton equations in classical mechanics, the microscopic equation determining quantum mechanical evolution, the Schr\"odinger equation, is time-reversible, i.e. invariant under time reversal. Therefore in quantum mechanics there exist the same problems as in classical mechanics~\cite{Davies} when one wants to deduce irreversible behaviour of thermodynamic systems.

One way to 'introduce' irreversibility in quantum mechanics is the treatment of open systems by means of master equations~\cite{Carmichael}~\cite{Weiss}. In this paper we consider another very general approach to this problem that does not need those specific assumption about an environment that is not part of the system considered. Instead, system \emph{and} environment are treated as a whole via pure Schr\"odinger dynamics; even for small systems (i.e. small accessible Hilbert-spaces) thermodynamic behaviour and quasi-irreversibility is shown to emerge for a large class of systems.

We have performed numerical simulations on relatively small quantum
systems (up to total state vector dimensions of $N^{tot}\approx
1000$). They are partitioned into a smaller ``part of interest'', the
'gas' $g$, and a larger part, the environment or 'container' $c$,
$N^{tot}= N^gN^c$. The interaction between those two subsystems has to
be small in comparison to the local energies
(i.e. $\erwwert{\hat{H}^{int}}\ll
\{\erwwert{\hat{H}^g},\erwwert{\hat{H}^c}\}$ for typical states). This
constraint as well as the partitioning is typical for all thermodynamic systems. Even if there is no energy exchanged between the two subsystems (microcanonical coupling), an environment ('container') is eventually needed to maintain the extensive parameters of the system considered like e.g. volume etc. As opposed to the open system approaches based on a bath, our environment $c$ does have neither a temperature nor a specific energy distribution. Its main function is to allow for entanglement between the two subsystems $c$ and $g$.

The numerical simulations shown in this paper are intended to test wether this entanglement may, indeed, be considered the origin of thermodynamics and irreversibility~\cite{Jochen1}~\cite{Jochen2}~\cite{Tasaki}.


\section{The model}
\label{sec_model}
The Hamiltonian of the full system will consist of the sum of the two local Hamiltonians of the gas and the container, respectively, plus the interaction Hamiltonian:
\begin{equation}
\hat{H}=\hat{H}^{(0)}+\hat{H}^{int}=\hat{H}^g\otimes \hat{1}^c + \hat{1}^g\otimes \hat{H}^c+\hat{H}^{int}.
\end{equation}
Here, $\hat{H}^g$ includes already the confining effect of the container by means of an effective potential; $\hat{H}^{int}$ describes dynamical corrections (see~\cite{Jochen2}).
The present analysis will focus on 'typical' Hamiltonians rather than on concrete physical models based on specifically interacting particles (as e.g. in~\cite{Saito}~\cite{Jensen}). Our system will thus be defined in terms of abstract energy spectra and statistical interactions. 

We consider the eigenbasis of $\hat{H}^{(0)}$, the
product-states $\ket{i}^g\otimes \ket{m,s}^c$ with $i=0,1,2,\hdots,
N^g-1$ denoting the non-degenerate energy eigenstates of $g$,
$m=0,1,2,\hdots$ the energy eigenstates of $c$ with corresponding
degeneracy $s=1,2,\hdots, n_m^c$. Any state vector can thus be written
as
\begin{equation}
\ket{\Psi} = \sum \Psi_{i;m,s} \ket{i}^g\ket{m,s}^c.
\end{equation}

The Hamiltonian models will be assumed to be defined with respect to
this basis. $\hat{H}^{(0)}$ is thus diagonal. For most of the simulations, the
simplest model for the gas is used: a two-level system with level
spacing $\Delta E^g$ defining the pertinent energy scale.

We assume that the product state basis is in no way a preferred basis
for the interaction $\hat{H}^{int}$. Its matrix is therefore treated
as a hermitian random matrix, specified by some distribution function
$w(H^{int}_{ij})$. If nothing was known, this distribution function
should not depend on the representation chosen, i.e. must be invariant
under any unitary transformation. This is guaranteed by taking
$w(H^{int}_{ij}) \propto \exp[ -\frac{1}{4} \Sp{(\hat{H}^{int})^2}]$
implying independent zero-mean Gaussian distributions with variance
$\sigma_0 =\sqrt{2}\Delta E^g\alpha$ for the diagonal matrix elements
and variance $\sigma = \Delta E^g\alpha$ for the real and imaginary
parts of the off-diagonal elements~\cite{Haake}. We induce weak coupling by requiring $\alpha \ll 1$, i.e. the energy scale of $\hat{H}^{int}$ is small compared with that of $\hat{H}^{(0)}$. Microcanonical constraints are built in by setting all blocks within $\hat{H}^{int}$ equal to zero, which would connect different energy states within $g$.


\section{Initial state and simulation}
To observe relaxation one has to choose rather special initial states,
since by far most of the states in the accessible region of total
Hilbert space are equilibrium states with respect to the smaller
subsystem $g$ \cite{Jochen2}~\cite{Tasaki}. In~\cite{Saito}
e.g. randomly chosen states are taken as initial states and no
relaxation can be observed. Here we choose product states of the two
subsystem states $\ket{\psi^g}$ and $\ket{\psi^c}$:
\begin{equation}
\ket{\Psi(t=0)}=\ket{\psi^g} \otimes \ket{\psi^c}.
\end{equation}
With this choice we also avoid accidentally taking an eigenstate of the total Hamiltonian $\hat{H}$ (including $\hat{H}^{int}$) which would lead to a plain oscillation but not to a decay.
We assume to know the corresponding energy distribution $p^{tot}(E^{tot})$, which is a constant of motion (closed total system). From $p^{tot}$ we can calculate the mean energy $U = \overline{E^{tot}}$.

The state vector $\ket{\Psi(t)}$ of the total system will be calculated numerically using a diagonalisation of the total Hamiltonian $\hat{H}$. Through $\hat{H}^{int}$ the two subsystems $g$ and $c$ entangle and can no longer be described by local state vectors $\ket{\psi^g}$ and $\ket{\psi^c}$. Local properties of the subsytems are described using the reduced density matrix of e.g. the gas:
\begin{equation}
\label{gl_Grundlagen_reduzierte_Dichtematrix}
\rho^g_{ij}=\sum_{m}\sum_{s}^{n^c_m}\bra{i;m,s}\hat{\rho}\ket{j;m,s},
\end{equation}
where $\hat{\rho}= \ket{\Psi}\bra{\Psi}$ is the density matrix of the total system. The diagonal elements of the reduced density matrix of the gas, $\rho^g_{ii}$, are the probabilities of finding the gas in the corresponding local state $\ket{i}^g$. The offdiagonal elements of $\hat{\rho}^g$ are a measure for local coherence.

From eq.~(\ref{gl_Grundlagen_reduzierte_Dichtematrix}), the reduced von-Neumann entropy $S^g$ of the gas can be calculated, a measure for the purity of the subsystem as well as for the entanglement between the two subsystems:
\begin{equation}
S^g(\hat{\rho}^g)=-k_B\Sp{\hat{\rho}^g\ln \hat{\rho}^g}
\label{eq_S_allg}
\end{equation}
(with the Boltzmann-constant $k_B$). $S^g$ is zero initially (product state) but not conserved.

In our simulations we set $\hbar = 1$ and $\Delta E^g = 1$ which gives time the unit of $\frac{\hbar}{\Delta E^g}$.


\section{Results}
\label{sec_results}

\subsection{Microcanonical coupling}
\label{subsec_mc}
\begin{figure}
\psfrag{WW}{\scriptsize Interaction}
\psfrag{delta_E}{$\Delta E^g$}
\psfrag{|0>}{$\ket{0}^g$}
\psfrag{|1>}{$\ket{1}^g$}
\psfrag{N}{$N^c$}
\resizebox{0.45\textwidth}{!}{\includegraphics{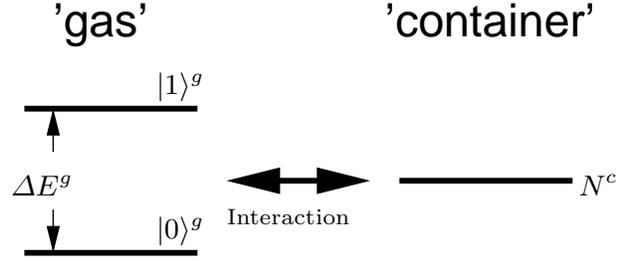}}
\caption{Pertinent relevant energy spectrum for the case of microcanonical coupling.}
\label{fig_spectrum_mc}
\end{figure}

\begin{figure}
\psfrag{S^g}[c]{\Huge $S^g$}
\psfrag{|rho^g_01|^2}[c]{\Huge \; \; $\sqrt{|\rho^g_{01}|^2}$}
\psfrag{Zeit}[c]{\Huge time}
\resizebox{0.45\textwidth}{!}{\includegraphics[angle=-90]{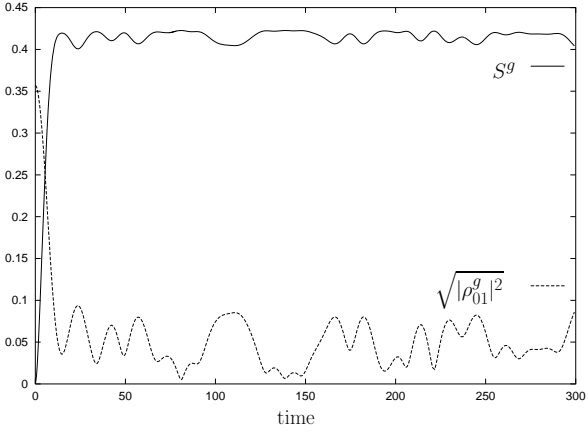}}
\caption{'2-level-gas' under microcanonical constraints: Von-Neumann entropy $S^g$ (in units of $k_B$) and absolute value of the off-diagonal element of the reduced density matrix as a function of time in units of $\frac{\hbar}{\Delta E^g}$. The container degeneracy is $N^c=50$.}
\label{fig_mc_1}
\end{figure}

 The microcanonical coupling in thermodynamics excludes energy exchange between the gas and the container, i.e. the energy distribution of the subsystems, $p^g$ and $p^c$, are separate constants of motion. Therefore, if we restrict ourselves to sharp energy initial states in the container, only one energylevel with degeneracy $N^c$ has to be considered in this subsystem. Fig.~\ref{fig_spectrum_mc} shows the energy spectrum for the case of microcanonical coupling. $E^g_0$ and $E^g_1$ are the energies of the two levels in the gas, $E^c$ of the container level. 
For a system as shown in fig.~\ref{fig_spectrum_mc} with $N^c=50$ ($i=0,1; m=0; s=1,2,\hdots N^c$), fig.~\ref{fig_mc_1} shows the behaviour of $\sqrt{|\rho^g_{01}|^2}$ (the off-diagonal element of the reduced density matrix of the 2-level-gas) and $S^g$ (its von-Neumann entropy).
In a system with microcanonical coupling, the diagonal elements of the reduced density matrix (here $\rho^g_{00}=0.15; \rho^g_{11}=0.85$) are constants of motion, but the off-diagonal element decreases from the value of the product state $\sqrt{0.15\cdot 0.85} = 0.3571$ towards zero, the value for a completely mixed state.
The average entropy in equilibrium of this simulation (after ten units of time) is (in units of $k_B$) $\overline{S^g}\approx 0.4157$, the maximal entropy $S^g_{max}\approx 0.4226$. The largest possible value for this setting would be that of a completely mixed state with $\rho^g_{01}=0$ and can be calculated using the eigenvalues of $\hat{\rho}^g$: $S^g_{max, th}\approx 0.4227$ which agrees with the results from the simulation. In all the simulations done, $\overline{S^g}$ is always smaller than $S^g_{max,th}$ and the difference between those two values decreases with increasing system size $N^c$.

\subsection{Canonical coupling}
\label{subsec_c}
\begin{figure}
\psfrag{WW}{\scriptsize Interaction}
\psfrag{delta_E}{$\Delta E^g$}
\psfrag{|0>}{$\ket{0}^g$}
\psfrag{|1>}{$\ket{1}^g$}
\psfrag{n^c_3}{$n^c_2$}
\psfrag{n^c_2}{$n^c_1$}
\psfrag{n^c_1}{$n^c_0$}
\resizebox{0.45\textwidth}{!}{\includegraphics{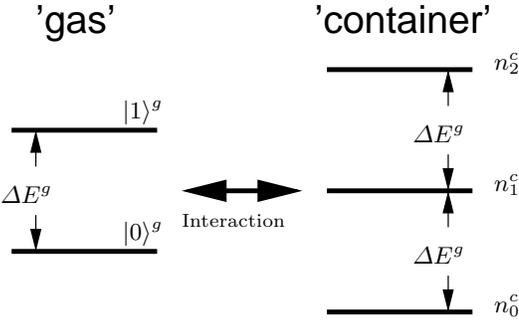}}
\caption{Pertinent relevant energy spectrum for the case of canonical coupling and sharp initial energies $E^c_1$ in the container. $n^c_j$ are the respective degeneracies.}
\label{fig_spectrum_c}
\end{figure}

We still restrict ourselves to a sharp initial energy in $c$, $E^c_1$. Conservation of the total energy distribution requires the consideration of only two more energies: $E^c_0=E^c_1-\Delta E^g$ and $E^c_2=E^c_1+\Delta E^g$. Fig.~\ref{fig_spectrum_c} shows this energy spectrum. Each of the three energylevels of the container $m=0,1,2$ is degenerate and contains $n^c_m$ levels.

The spectrum of the container in between those three levels does not
play any role. Important, however, are the corresponding
degeneracies. Here the usual assumptions about large systems are
made~\cite{Brenig}, i.e. the degeneracy of the energy levels is taken
to grow exponentially with the energy: $n^c(E^c)\propto \ee^{\gamma
E^c}$, which we call the thermodynamic scheme of degeneracy.

\begin{figure}
\psfrag{Sim1}[c]{\Huge $\scriptstyle \alpha = 0.005$ \;}
\psfrag{Sim2}[c]{\Huge $\scriptstyle \alpha = 0.005$\; }
\psfrag{Sim3}[c]{\Huge $\scriptstyle \alpha = 0.001$ \;}
\psfrag{rho^g_00}{\Huge $\rho^g_{00}$}
\psfrag{Zeit}[c]{\Huge time}
\resizebox{0.45\textwidth}{!}{\includegraphics[angle=-90]{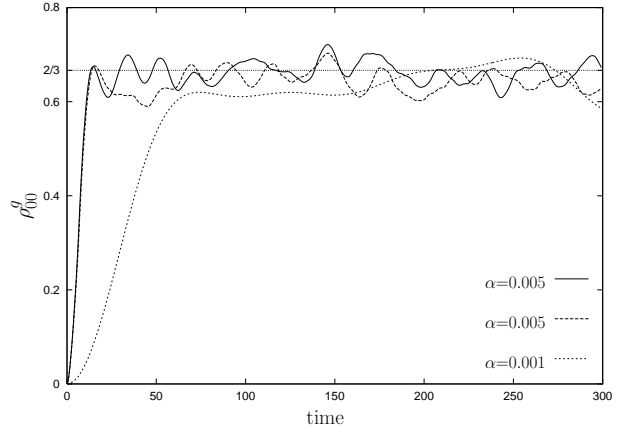}}
\caption{'2-level-gas' under canonical constraints: Simulations for three different interactions and the same initial state in $g$.}
\label{fig_results_c_1}
\end{figure}

Fig.~\ref{fig_results_c_1} shows the probability $\rho^g_{00}$ of
finding the subsystem $g$ in the ground state $\ket{0}^g$ plotted over
time for three different simulations that all had the same energy
spectrum: $E^g_0=0, E^g_1=1$ (both non-degenerate); $E^c_0=0;
n^c_0(E^c_0)=50; E^c_1=1; n^c_1(E^c_1)=100; E^c_2=2;
n^c_2(E^c_2)=200$. The first two simulations with both $\alpha =
0.005$ but different random numbers in $\hat{H}^{int}$ show fast
relaxation. The smaller the entries in the interaction-Hamiltonian,
the slower equilibrium will be reached which is shown in the third simulation using $\alpha = 0.001$. The initial states for these simulations are $\ket{\Psi(t=0)} = \ket{1}^g \otimes \ket{n}^c$ with $n^c_0 \le n < n^c_0+n^c_1$.

The equilibrium value for this system may be calculated using Boltzmann's a priori postulate applied to the energy eigenstates of the total quantum system:
\begin{equation}
p^g(E^g)=\frac{n^c(U-E^g)n^g(E^g)}{n^{tot}(U)}
\label{eq_prob_dist};
\end{equation}
\begin{equation}
\rho^g_{00}\equiv p^g(E^g=0)=\frac{n^c(2-0)n^g(0)}{n^{tot}(2)}=\frac{200\cdot 1}{200+100}=\frac{2}{3}
\end{equation}
which is in surprisingly good accord with our simulations; the latter also show the important fact of independence of the actual microscopic details of the interaction, as required for analogies to thermodynamic behaviour. This fact does not depend on any assumptions about the structure of degeneracy in the system but is a general aspect of composite quantum systems. We have checked and confirmed the independence of the equilibrium value from the starting level $n$ in the container. Furthermore, neither the equilibrium value nor its variance depend sensitively on $\alpha$, the parameter for the strength of the coupling, as long as $\alpha\ll 1$. We have seen deviations from thermodynamic behaviour for some large $\alpha$, but - without supporting a clear-cut message - this parameter-space has not been included in our present investigation.

\begin{figure}
\psfrag{|a_0|^2=0.9}[c]{\Huge $\scriptstyle p_0(\ket{0}^g)=0.9$}
\psfrag{|a_0|^2=0.5}[c]{\Huge $\scriptstyle p_0(\ket{0}^g)=0.5$}
\psfrag{|a_0|^2=0}[c]{\Huge $\scriptstyle p_0(\ket{0}^g)=0$}
\psfrag{Zeit}[c]{\Huge time}
\psfrag{rho^g_00}{\Huge $\rho^g_{00}$}
\resizebox{0.45\textwidth}{!}{\includegraphics[angle=-90]{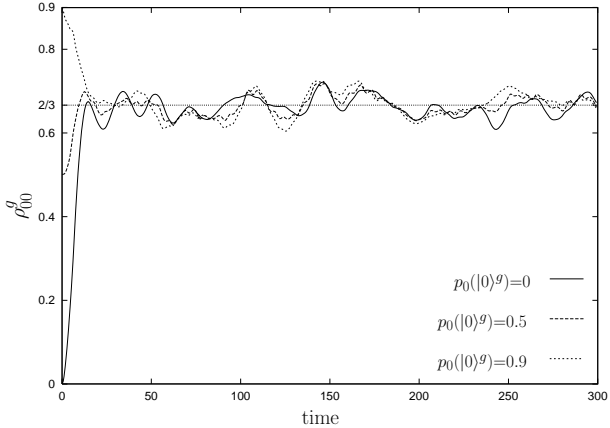}}
\caption{'2-level-gas' as of fig.~\ref{fig_results_c_1}: Simulations for three different initial probabilities $p_0(\ket{0}^g)$.}
\label{fig_results_c_2}
\end{figure}

However, the full independence of the initial conditions, i.e. of the initial energy distribution in the subsystem $g$ can only be reached for 'thermodynamic schemes' of degeneracy (defined above), as can easily be shown. In Fig.~\ref{fig_results_c_2} three different starting probabilities $0, 0.5$ and $0.9$ in state $\ket{0}^g$ lead to the same equilibrium value of $\frac{2}{3}$.

To emphasize the distinction between our model and common 'system-bath' models or open system models~\cite{Weiss}, we calculate the occupation of the container levels in these simulations. Whereas the latter models take for granted a thermal (Boltzmann) distribution of the energy in the bath, we obtain (i) different distributions for different initial conditions (but same distributions in $g$) and (ii) distributions that are far from any thermal distribution (e.g. for the case of $p_0(\ket{0}^g)=0.5$ the equilibrium occupation of the degenerate container energy state with $E^c_1=1$ is larger than the occupation of the lower state with $E^c_0=0$ by a factor of 3).

\begin{figure}
\psfrag{S^g}[c]{\Huge $S^g[k_B]$}
\psfrag{|a_0|^2=0.9}[c]{\Huge $\scriptstyle p_0(\ket{0}^g)=0.9$}
\psfrag{|a_0|^2=0.5}[c]{\Huge $\scriptstyle p_0(\ket{0}^g)=0.5$}
\psfrag{|a_0|^2=0}[c]{\Huge $\scriptstyle p_0(\ket{0}^g)=0$}
\psfrag{Zeit}[c]{\Huge time}
\resizebox{0.45\textwidth}{!}{\includegraphics[angle=-90]{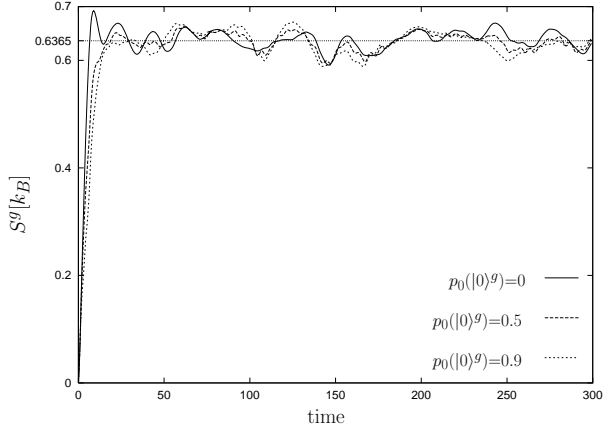}}
\caption{'2-level-gas' as of fig.~\ref{fig_results_c_2}: The von-Neumann entropy for the simulations from fig.~\ref{fig_results_c_2}.}
\label{fig_results_c_3}
\end{figure}

For a closer connection with thermodynamics, the von-Neumann entropy of the subsystem $g$ can be calculated according to eq.~(\ref{eq_S_allg}). This has been done for the three simulations from fig.~\ref{fig_results_c_2} and is shown in fig.~\ref{fig_results_c_3}. As expected for thermodynamic systems, the entropy increases up to an equilibrium value that does not depend on the initial conditions and fluctuates around this value (cf.~\cite{Wang}). The equilibrium value from the simulations is in accord with the maximal possible value for a $2\times 2$ matrix $\hat{\rho}^g$ with the equilibrium values of $\rho^g_{00}$ and $\rho^g_{11}$ on the diagonal and zeros in the offdiagonal (eq.~(\ref{eq_S_allg})): $S^g_{eq,th}\approx 0.6365k_B$.

\begin{figure}
\psfrag{simulations}[c]{\Huge simulations}
\psfrag{lsqf}[r]{\Huge least square fit}
\psfrag{(Delta rho)^2}{\Huge $\left( \Delta \rho^g_{00}\right)^2$}
\psfrag{n^c(E^c_2)}[t]{\Huge $n^c(E^c_1)$}
\resizebox{0.45\textwidth}{!}{\includegraphics[angle=-90]{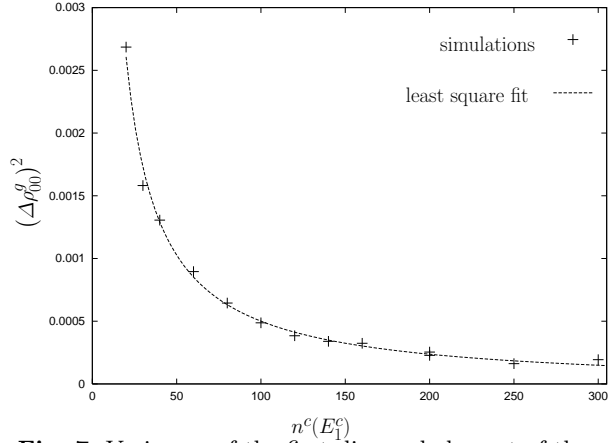}}
\caption{Variances of the first diagonal element of the reduced density matrix of the gas over system size for 13 simulations plus a least square fit (line).}
\label{fig_results_c_var}
\end{figure}

All the local measures considered so far exhibit fluctuations around their equilibrium values after relaxation. Fig.~\ref{fig_results_c_var} shows the variances of $\rho^g_{00}$ of 13 simulations that all produced the same equilibrium value of approximately $\frac{2}{3}$. The variances are plotted over the level of degeneracy $n^c(E^c_1)$ of the second (starting) energy level of the container as a measure for system size.
The line in fig.~\ref{fig_results_c_var} is the least square fit with $(\Delta \rho^g_{00})^2\approx \frac{0.053}{n^c(E^c_1)}$, i.e. the variance decreases with increasing system size. This is the result from standard thermodynamics~\cite{Brenig} and also follows analytically for the quantum case of bi-partite systems~\cite{Jochen_pers}.

\begin{figure}
\psfrag{delta_E}{$\Delta E^g$}
\psfrag{n^c_1}{$n^c_0$}
\psfrag{n^c_2}{$n^c_1$}
\psfrag{n^c_3}{$n^c_2$}
\psfrag{n^c_4}{$n^c_3$}
\psfrag{n^c_5}{$n^c_4$}
\psfrag{|0>^g}{$\ket{0}^g$}
\psfrag{|1>^g}{$\ket{1}^g$}
\psfrag{|2>^g}{$\ket{2}^g$}
\psfrag{|3>^g}{$\ket{3}^g$}
\psfrag{|4>^g}{$\ket{4}^g$}
\psfrag{E^g_2}[lb]{$E^g$}
\psfrag{WW}{\scriptsize Interaction}
\psfrag{E^g}[l]{$E^g$}
\psfrag{1}{0}
\psfrag{2}{1}
\psfrag{3}{2}
\psfrag{4}{3}
\psfrag{5}{4}
\resizebox{0.45\textwidth}{!}{\includegraphics{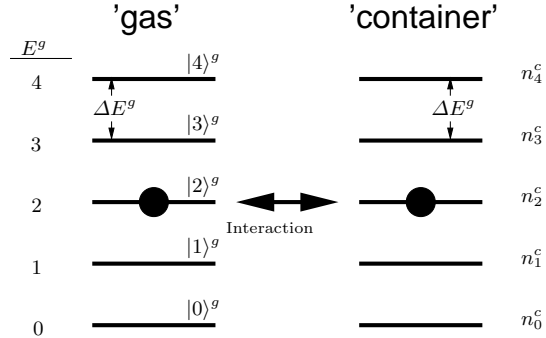}}
\caption{Energy spectrum used for the two simulations reproducing the Boltzmann-distribution. The two dots symbolize the sharp initial energies in both the container and the gas.}
\label{fig_spectrum_c_2}
\end{figure}

To further clarify the connection between quantum mechanics of small
embedded systems and thermodynamic behaviour, the number of levels in
the gas-subsystem has finally been increased from two to five. We
restrict ourselves from now on to sharp initial energies, not only in $c$, but also in $g$, which reduces the number of energies that
have to be considered for the dynamics to five in the container (Fig.~\ref{fig_spectrum_c_2}). Again, it does not matter, how the spectrum looks like in between those levels: The conservation of energy distribution of the Schr\"odinger equation requires only those levels shown here.

Two simulations have been carried out based on the two different degeneracy schemes shown in table~\ref{tab_c_Boltzmann}.

\begin{figure}
\psfrag{simulation 1}[c]{\Huge  simulation 1 \;}
\psfrag{simulation 2}[c]{\Huge  simulation 2 \;}
\psfrag{least square fit 1}[c]{\Huge  least square fit 1 \; \; \;}
\psfrag{least square fit 2}[c]{\Huge  least square fit 2 \; \; \;}
\psfrag{p^g}{\Huge $\overline{p^g}$}
\psfrag{E^g}[c]{\Huge $E^g$}
\resizebox{0.45\textwidth}{!}{\includegraphics[angle=-90]{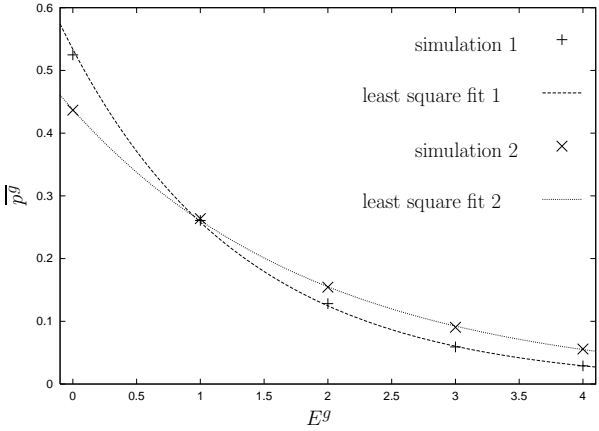}}
\caption{'5-level-gas' under canonical constraints: Averages of the probabilities to find the subsystem $g$ in the corresponding energy level for the two simulations together with the least square fits.}
\label{fig_results_Boltzmann}
\end{figure}

The degrees of degeneracy obey our 'thermodynamic condition' as defined above: for simulation 1, $n^c(E^c)=6\cdot 2^{E^c}$, for simulation 2, $n^c(E^c)\approx 17\cdot (1.7)^{E^c}$.
\begin{table}
\caption{The two degeneracy schemes used to reproduce the Boltzmann-distribution.}
\label{tab_c_Boltzmann}
\begin{tabular}{c|ccccc}
\hline\noalign{\smallskip}
$E^c$ & 0 & 1 & 2 & 3 & 4 \\
\noalign{\smallskip}\hline\noalign{\smallskip}
$n^c(E^c)$ \, for sim.1 & 6 & 12 & 24 & 48 & 96 \\
$n^c(E^c)$ \, for sim.2 & 17 & 29 & 49 & 84 & 142 \\
\noalign{\smallskip}\hline
\end{tabular}
\end{table}
For both simulations averages of the diagonal elements of the reduced density matrices are calculated after relaxation and are shown in fig.~\ref{fig_results_Boltzmann}.

The energy distributions in the gas follow a Boltzmann-distribution, the expected distribution for a thermodynamic system under canonical coupling. The least square fits give $p^g(E^g)\approx 0.53\cdot \ee^{-0.73E^g} \approx 0.53\cdot (2.1)^{-E^g}$ for simulation 1 and $p^g(E^g)\approx 0.44\cdot \ee^{-0.52E^g}\approx 0.44\cdot (1.68)^{-E^g}$ for simulation 2.

Through the exponents of those two fits, two different temperatures can be introduced to describe the two different systems. The value of those temperatures in our model is determined completely by the degeneracy scheme of the larger subsystem $c$. (Under more general conditions the temperature will also depend on the initial energy distribution.) Again, we did not 'produce' this result by enforcing a thermal energy distribution of adequate temperature into the container. Only the smaller of the two subsystems evolves into a thermal equilibrium state, the larger one does not.


\section{Summary}
\label{sec_summary}
We have studied small quantum systems, in which a partition into two differently sized subsystems typically leads to a quasi-irreversible behaviour in the smaller subsystem (the 'gas') that can be called thermodynamic in many respects, even though the whole system is described by a pure Schr\"odinger-dynamics. For our specifically chosen initial states the 'container' does \emph{not} approach a thermal equilibrium state. As opposed to master equation treatments that assume no entanglement at all between the considered system and its environment, the only reason for one subsystem to approach equilibrium \emph{is} the entanglement between the two subsystems.

All the results obtained here are in good accordance with recent attempts to base thermodynamics on the microscopic principles of quantum mechanics~\cite{Jochen1}~\cite{Jochen2}~\cite{Tasaki} and are to be seen as a generaliziation of real physical models described before~\cite{Saito}~\cite{Jensen}.

Our paper may also shed new light on the recent debate about
possibilities to violate thermodynamic expectations, in particular the
second law of thermodynamics~\cite{Opatrny}~\cite{Capek}. The
conditions found in our model for the smaller subsystem to behave
thermodynamically can, if broken,  also be used as a gateway to
non-thermodynamic systems. E.g. a non-exponential dependence of the
degeneracy of the energy would lead to energy distributions within the
'gas' that were no longer the Boltzmann-type. Furthermore, for small
systems the exponential decay (in terms of pertinent occupation
probabilities) may change towards what would look more like a
Gaussian. Further investigations are in progress.

Financial support by the Deutsche Forschungsgemeinschaft is gratefully acknowledged.


\end{document}